\documentclass[aps,prb,nobibnotes,twocolumn,superscriptaddress,shortbibliography]{revtex4-2}

\usepackage{amsfonts}
\usepackage{mathrsfs}
\usepackage{amsmath}
\usepackage{color}
\usepackage{graphicx}
\usepackage{bm}
\usepackage{amssymb}
\usepackage{xspace}
\usepackage{epstopdf}
\usepackage{dcolumn}
\usepackage{longtable}
\usepackage{multirow}
\usepackage{float}
\usepackage{comment}
\usepackage{lineno}

\usepackage[colorlinks=true, letterpaper=true, pdfstartview=FitV,  linkcolor=blue, citecolor=blue, urlcolor=blue]{hyperref}

\makeatother

\begin{document}

\title{Landau level spectrum and magneto-optical conductivity in tilted Weyl semimetal}

\author{Pu Liu}
\affiliation{Centre for Quantum Physics, Key Laboratory of Advanced Optoelectronic Quantum Architecture and Measurement (MOE), School of Physics, Beijing Institute of Technology, Beijing, 100081, China}
\affiliation{Beijing Key Lab of Nanophotonics \& Ultrafine Optoelectronic Systems, School of Physics, Beijing Institute of Technology, Beijing, 100081, China}

\author{Chaoxi Cui}
\affiliation{Centre for Quantum Physics, Key Laboratory of Advanced Optoelectronic Quantum Architecture and Measurement (MOE), School of Physics, Beijing Institute of Technology, Beijing, 100081, China}
\affiliation{Beijing Key Lab of Nanophotonics \& Ultrafine Optoelectronic Systems, School of Physics, Beijing Institute of Technology, Beijing, 100081, China}

\author{Xiao-Ping Li}
\email{xpli@imu.edu.cn}
\affiliation{School of Physical Science and Technology, Inner Mongolia University, Hohhot 010021, China}

\author{Zhi-Ming Yu}
\email{zhiming\_yu@bit.edu.cn}
\affiliation{Centre for Quantum Physics, Key Laboratory of Advanced Optoelectronic Quantum Architecture and Measurement (MOE), School of Physics, Beijing Institute of Technology, Beijing, 100081, China}
\affiliation{Beijing Key Lab of Nanophotonics \& Ultrafine Optoelectronic Systems, School of Physics, Beijing Institute of Technology, Beijing, 100081, China}
\author{Yugui Yao}
\affiliation{Centre for Quantum Physics, Key Laboratory of Advanced Optoelectronic Quantum Architecture and Measurement (MOE), School of Physics, Beijing Institute of Technology, Beijing, 100081, China}
\affiliation{Beijing Key Lab of Nanophotonics \& Ultrafine Optoelectronic Systems, School of Physics, Beijing Institute of Technology, Beijing, 100081, China}

\begin{abstract}
We present a systematic investigation of the magnetoresponses of the Weyl points (WPs) with a topological charge of $n=2, 3$ and $4$, and with both linear and quadratic energy tilt. The linear tilt always tends to squeeze the Landau levels (LLs) of both conduction and valence bands of all the WPs, and eventually leads to LL collapse in the type-II phase. However, the quadratic energy tilt has more complex influences on the LLs of the unconventional WPs. For charge-$n$ ($n=2, 4$) WP, the influence of the quadratic tilt on the LLs of conduction and valence bands are opposite, i.e. if the LLs of conduction (valence) bands are squeezed, then that of the valence (conduction) bands are broadened, and the squeezed LL spectrum will be collapsed in type-III phase. But, the LL collapse generally can not be found in the type-III charge-3 WP. Moreover, for charge-$n$ ($n=2, 3$) WP, the quadratic tilt breaks the degeneracy of the chiral LLs regardless of the direction of the magnetic field, leading to additional optical transitions and magneto-optical conductivity peaks at low frequencies. Interestingly, the four chiral LLs in charge-$4$ WP are always not degenerate. Hence, there inevitably exist magneto-optical conductivity peaks at low frequencies for charge-$4$ WP. Since the density of state of the LL spectrum is very large, one can expect that the low-frequency magneto-optical responses in unconventional WPs would be significant and may be used for developing efficient terahertz photodetectors.
\end{abstract}

\maketitle
\section{Introduction}

Weyl semimetal, a novel topological state of matter that possess momentum-space
singularities, has been attracting broad interests in current research \cite{NPArmitage,XiaoliangQi}. Such singularity, named as Weyl point (WP),
is a kind of double degeneracy formed by non-degenerate conduction
and valence bands \cite{XWan}. Around the degeneracy, the conventional WP exhibits
linear dispersion along any direction in momentum space, respecting the Weyl equation proposed in high-energy physics \cite{XWan}. Hence,
each conventional WP is assigned a chirality (unit topological charge)
with $|{\cal C}|=1$.

However, the crystals do not have Lorentz symmetry, and hence the WP in crystals can take different forms, generally manifested in two aspects.
The first one is that with certain symmetries such as rotation axis,
multiple conventional Weyl points with same topological charge can
merge together, leading to unconventional WPs \cite{GXu,CFang,ZMYu}.
Recently, we show that there  are only three possibilities for the
crystalline-symmetry protected unconventional WPs \cite{ZMYu,GuiBinLiu,ZeyingZhang}, which
have topological charge $|{\cal C}|=2,\ 3$ and $4$, respectively.
The charge-$2$ (C-2) and C-3 WPs exhibit linear dispersion along
rotation axis, and nonlinear dispersion in the plane normal to rotation
axis \cite{CFang}. More interestingly, the C-4 WP features nonlinear dispersion
along any direction in momentum space \cite{ZMYu,TZhang,QBLiu}.
The second aspect is the tilt of the Weyl cone, as many energy tilt terms are compatible with the crystalline symmetry.
For conventional C-1 WP, only linear energy term is relevant, as the leading order of $\boldsymbol{k}$ in the
corresponding Hamiltonian is linear. But for C-2 and C-3 WPs, both
linear and quadratic energy tilt terms are relevant and symmetry allowed \cite{CFang,ZMYu}.
Because the C-4 WP is located at the high-symmetry point with time-reversal
symmetry ${\cal T}$ or ${\cal T}^{\prime}=\{{\cal T}|t_{0}\}$ with
$t_{0}$ a half lattice translation \cite{ZMYu,GuiBinLiu,ZeyingZhang}, which forbids the
linear energy tilt, hence it has only quadratic energy tilt.

Both linear and quadratic energy tilt do not change the essential topological properties  of the WPs, such as  topological charge, but they do have
important influence on the geometry of the Fermi surface. By increasing
the linear energy tilt, the C-$n$ ($n=1,2,3$) WP will undergo a
phase transition form type-I WP to type-II WP \cite{AASoluyanov}, for which the Fermi
surface respectively are a point and a surface constructed by electron
pocket and hole pocket \cite{AASoluyanov,ZhiMingYu2,TEBrien,KeDeng,MYYao,Lunan}. Similarly, the quadratic energy tilt also
can transform the Fermi surface of C-$n$ ($n=2,3,4$) WP from a point to a surface, but which  here is constructed by two electron
or hole pockets, distinct from the case in type-II WP  \cite{XiaopingLi,XiaopingLi2}.
Such WPs hence are referred to as type-III to be distinguished from the type-II ones.
Since most of the material properties are determined by the geometry
of Fermi surface and the density of states around Fermi level, one
can expect that the WPs with different topological charge or with
and without energy tilt will exhibit completely different signatures,
which is also an important basis for experimentally identifying these
topological states.

Magneto-optical conductivity can  serve as an experimental tool for observing salient features of systems, which are not available in
direct-current measurement \cite{YuxuanJiang,YHayashi,AAkrap}. For example, the information like energy
gap and Fermi velocity of the band structure of systems can be extracted
from these resonant peaks of the optical conductivity \cite{BXu,PECAshby,SeongjinAhn,DNeubauer,Lunan,CJTabert,CJTabert2,YuxuanJiang,YHayashi}. Hence, optical
spectrum measurement is a standard and fundamental approach to detect and understand the band structure of target materials. Recently, the
magneto-optical conductivity of topological Weyl semimetals has been studied by several  works \cite{MUdagawa1,PECAshby2,JDMalcolm,JMShao,YongSun,Marcus}.
Ashby and Carbotte studied the magneto-optical conductivity of the C-1 WP without energy tilt  for different Fermi levels and scattering
rates \cite{PECAshby2}. By introducing linear energy tilt, the C-1 WP becomes anisotropic
and the LLs of system has a strong dependence on the direction of $B$ field \cite{ZhiMingYu2}.
Particularly, in type-II phase, the LLs of C-1 WP would be collapsed at a critical angle between the $B$ field and the tilt, and the magneto-optical conductivity is distinguished from that of type-I C-1 WP \cite{ZhiMingYu2,MUdagawa1,Serguei}. The magneto-optical
conductivity of C-2 and C-3 WPs with and without linear energy tilt
were also investigated in Ref. {\cite{Marcus,YJJIN,Serguei}}.
However, the magneto-response of recently proposed C-4 WP and that of the C-$n$ ($n=2,3$) WPs with quadratic energy tilt has yet not been explored.

In this work, we present a systematic investigation of the magneto-response
of unconventional WPs with both linear and quadratic energy tilt,
based on the low-energy effective Hamiltonian and Kubo formula. We
find that for C-2 and C-4 WPs, the additional quadratic energy tilt
tends to squeeze the LLs of one of the valence and conduction band,
and broadens the LLs of other band. Hence, when quadratic energy tilt
is large and the C-2 (C-4) WP becomes type-III one, only valence or
conduction band features LL collapse, which is sharp contrast to the
type-II WPs, where the LLs of both valence and conduction bands collapse \cite{ZhiMingYu2}.
Interestingly, the LL collapse generally can not be realized in C-3 WP
regardless of the presence or absence of the quadratic energy tilt.

The quadratic energy tilt also has important influence on the magneto-optical
conductivity of the unconventional WPs. In the absence of quadratic
energy tilt, both C-2 and C-3 WPs have degenerate chiral LLs when the $B$ field is parallel to the principal rotation axis of systems \cite{YongSun,YangGao}.
Such degeneracy will be broken by the quadratic energy tilt, which
make a new transition between the chiral LLs and lead to additional
peak(s) in the magneto-optical conductivity spectrum at low frequencies.
At the same time, the original magneto-optical conductivity peaks
will be splitted. But, the four chiral LLs in C-4 Weyl point is always
not degenerate \cite{ZMYu,ChaoxiCui} and the absorption peaks from the transition between
different chiral LLs generally  are much significant than that in C-2 and C-3
WPs. Besides, similar to the previous investigation on type-II WPs \cite{ZhiMingYu2},
the type-III WPs exhibit many intraband absorption peaks at low frequencies.

This paper is organized as follows. In Sec. \ref{sec:II}, we analytically calculate the LLs for C-2 and C-3 WPs with both linear and quadratic energy tilt when the magnetic field along the $z$ direction and numerically calculate the LL spectrum of C-$n$ ($n=2, 3, 4$) WPs with different tilt parameters. Beside, we discuss the LL collapse in the unconventional WPs  based on  a semiclassical picture.
Then, in Sec. \ref{sec:III}, we present the magneto-optical transition selection rules and show the numerical results for the corresponding magneto-optical conductivity for C-$n$ ($n=2,3,4$). At last, we summarize our results in Sec. \ref{sec:IV}.

\section{LL Spectrum \label{sec:II}}

\subsection{C-2 and C-3 WPs }

\begin{figure*}
\includegraphics[width=16 cm]{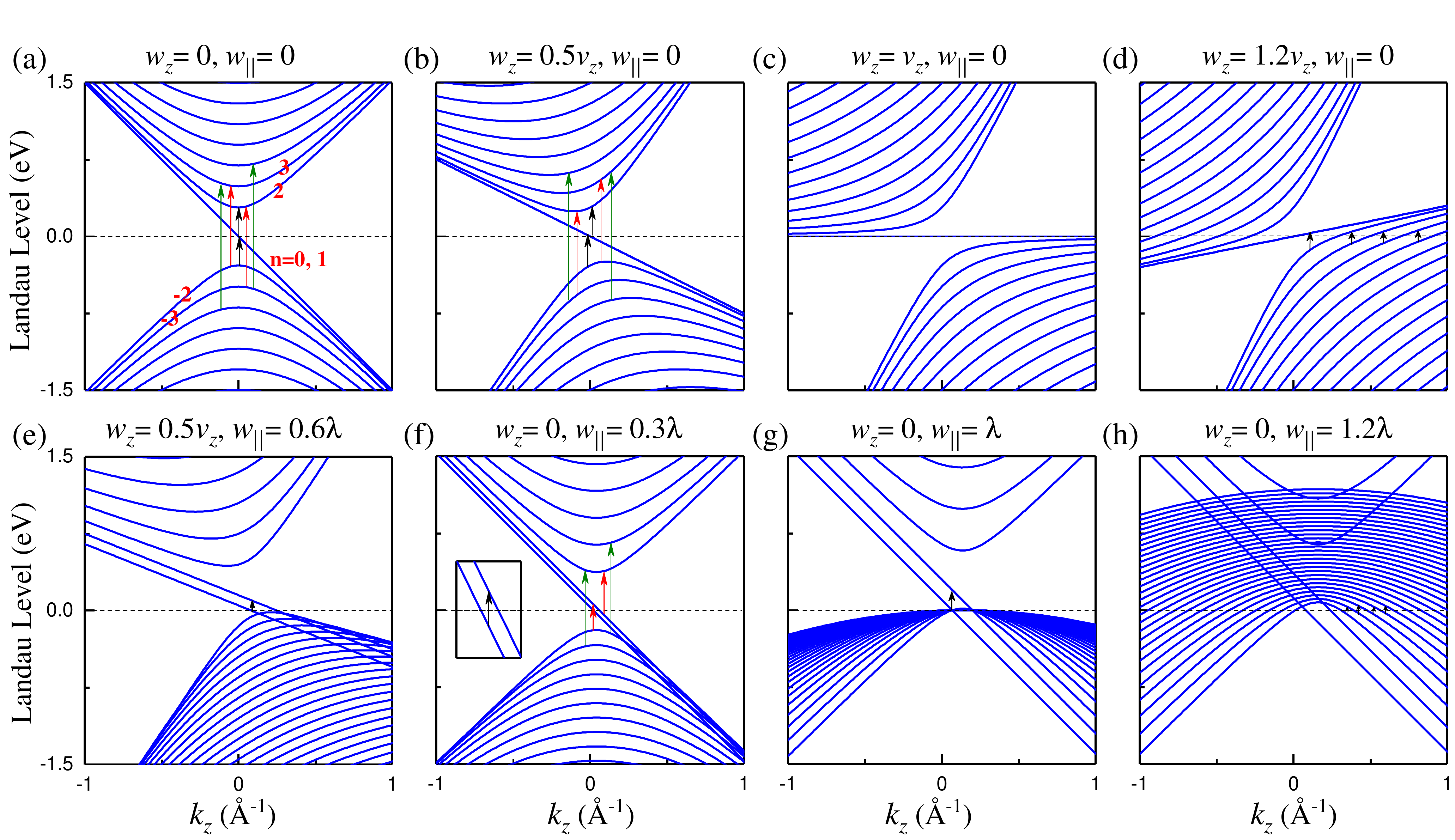}
\caption
{The LL spectrum of the C-2 WPs along $k_z$ with different energy tilts. We set $v_z$=1.5 eV$\cdot$\AA, $\lambda$=0.1 eV$\cdot$\AA$^2$ and $l_B=1$ \AA.
The arrows  mark some representative optical transitions.}
\label{fig.1}
\end{figure*}

A general low-energy effective Hamiltonian of C-2 (C-3) WP without
energy tilt can be written as \cite{ZMYu}
\begin{equation}
{\cal H}_{0}=v_{z}k_{z}\sigma_{z}+\lambda(k_{-}^{m}\sigma_{+}+k_{+}^{m}\sigma_{-}),\label{eq:ham1}
\end{equation}
where $v_{z}$ and $\lambda$ are real model parameters, $m=2$ ($m=3$),
$\sigma_{\pm}=(\sigma_{x}\pm i\sigma_{y})/2$ with $\sigma_{i}$ ($i=x,y,z$)
the Pauli matrix and $k_{\pm}=k_{x}\pm ik_{y}$. One observes that
the leading order of $\boldsymbol{k}$ in the effective Hamiltonian
(\ref{eq:ham1}) along $k_{z}$-direction is linear and that in $k_{x}$-$k_{y}$
plane is quadratic (cubic) for $m=2$ ($m=3$).
Hence, a linear energy tilt term $\propto k_{i}$ will compete with $v_{z}k_{z}\sigma_{z}$ along $k_{z}$-direction and dominates the energy dispersion around the
WP in $k_{x}$-$k_{y}$ plane. In practice, however, the C-2 and C-3
WPs are located at high-symmetry line ($k_{z}$ axis here) with multiple-fold
rotation symmetry, which forbids the appearance of linear and cubic terms in the $k_{x}$-$k_{y}$ plane  \cite{ZMYu}. Consequently, the relevant energy tilt
for C-2 (C-3) WP includes two parts: a linear energy tilt along $k_{z}$-direction
and a quadratic energy tilt in the $k_{x}$-$k_{y}$ plane, for which
the effective Hamiltonians read
\begin{equation}
{\cal H}_{tilt}=w_{z}k_{z}+w_{||}k_{||}^{2},\label{eq:tilt1}
\end{equation}
with $k_{\parallel}=\sqrt{k_{x}^{2}+k_{y}^{2}}$. The energy dispersion
of the C-$m$ ($m=2, 3$) WP with energy tilt (\ref{eq:tilt1}) is
\begin{eqnarray}
E_{\pm} & = & w_{z}k_{z}+w_{||}k_{||}^{2}\pm\sqrt{v_{z}^{2}k_{z}^{2}+\lambda^{2}k_{\parallel}^{2m}}.\label{eq:hamB}
\end{eqnarray}
When $|w_{z}|>|v_{z}|$, the Weyl cone is overtilted, leading to a
type-II C-2 (C-3) WP. But the influence of quadratic energy tilt on
C-2 and C-3 WP is different, which then leads to different consequences
for the LL spectrum, as shown below. For C-2 WP, it becomes
a type-III WP when $|w_{||}|>|\lambda|$, as its band structure in
$k_{x}$-$k_{y}$ plane is $E_{\pm}=(w_{||}\pm\lambda)k_{||}^{2}$.
For C-3 WP  with energy tilt, its energy dispersion in $k_{x}$-$k_{y}$
plane is $E_{\pm}=w_{||}k_{||}^{2}\pm\lambda k_{\parallel}^{3}$,
indicating  the band structure of C-3 WP is dominated by $w_{||}k_{||}^{2}$ for small $k_{\parallel}$. Hence, an arbitrary small $|\lambda|$
can transform a type-I C-3 WP  to a type-III one.
But, it should be noticed that for large $k_{\parallel}$, the band structure of C-3 WP  is dominated by $\pm\lambda k_{\parallel}^{3}$.

We first consider a uniform magnetic field along $z$ direction,
as in such case the LL spectrum of C-2 (C-3) WP can be analytically
obtained. We make the usual Peierls substitution $\boldsymbol{k}\rightarrow\boldsymbol{\Pi}=\boldsymbol{k}+e\boldsymbol{A}/\hbar$
in the effective Hamiltonian with the vector potential $\boldsymbol{A}=(0,Bx,0)$.
Here, we have chosen the Landau gauge. By introducing the creation
and annihilation operators
\begin{eqnarray}
\hat{a}=\frac{l_{B}}{\sqrt{2}\hbar}(\Pi_{x}+i\Pi_{y}), & \ \  & \hat{a}^{\dagger}=\frac{l_{B}}{\sqrt{2}\hbar}(\Pi_{x}-i\Pi_{y}),
\end{eqnarray}
 the Hamiltonian ${\cal H}={\cal H}_{0}+{\cal H}_{tilt}$ can be rewritten
as
\begin{eqnarray}
H&=&w_{z}k_{z}+v_{z}k_{z}\sigma_{z}+\omega_{c}(2\hat{a}^{\dagger}\hat{a}+1)\nonumber\\
& &+\lambda\left(\frac{\sqrt{2}}{l_{B}}\right)^{m}\left[\hat{a}^{m}\sigma_{+}+\left(\hat{a}^{\dagger}\right)^{m}\sigma_{-}\right],
\end{eqnarray}
where $l_{B}$=$\hbar/\sqrt{eB}$ denotes the magnetic length and $\omega_c=w_{||}/l_{B}^{2}$.

After straightforward calculations, the LL spectrum and the eigenstates of C-2 WP ($m=2$) are obtained as
\begin{eqnarray}
\varepsilon_{n}(k_{z}) & = & (2|n|+1)\omega_{c}+\left(w_{z}-v_{z}\right)k_{z}, \label{eq:LL20} \\
\Psi_{n} & = & (0,\left\vert n\right\rangle)^T,\label{eq:wf20}
\end{eqnarray}
for $n=0,1$, and
\begin{eqnarray}
\varepsilon_{n}(k_{z}) & = & (2|n|-1)\omega_{c}+w_{z}k_{z}\pm\Gamma_{n}(k_{z}),\label{eq:LL21}\\
\Psi_{n} & = & (\alpha_{n}\left\vert n-2\right\rangle
,\beta_{n}\left\vert n\right\rangle )^T,\label{eq:wf21}
\end{eqnarray}
for $|n|\geqslant2$ with
\begin{eqnarray}
\Gamma(k_{z}) & = & \sqrt{\left(v_{z}k_{z}-2\omega_{c}\right)^{2}+\frac{4\lambda^{2}}{l_{B}^{4}}|n|(|n|-1)}, \\
\alpha_n&=&\sqrt{\frac{v_zk_z-2\omega_{c}\pm \Gamma(k_z)}{\pm2\Gamma(k_z)}}, \\
\beta_n&=&\sqrt{\frac{-v_zk_z+2\omega_{c}\pm \Gamma(k_z)}{\pm2\Gamma(k_z)}},
\end{eqnarray}
and $|n\rangle$  the harmonic oscillator eigenstate.

\begin{figure*}
\includegraphics[width=16 cm]{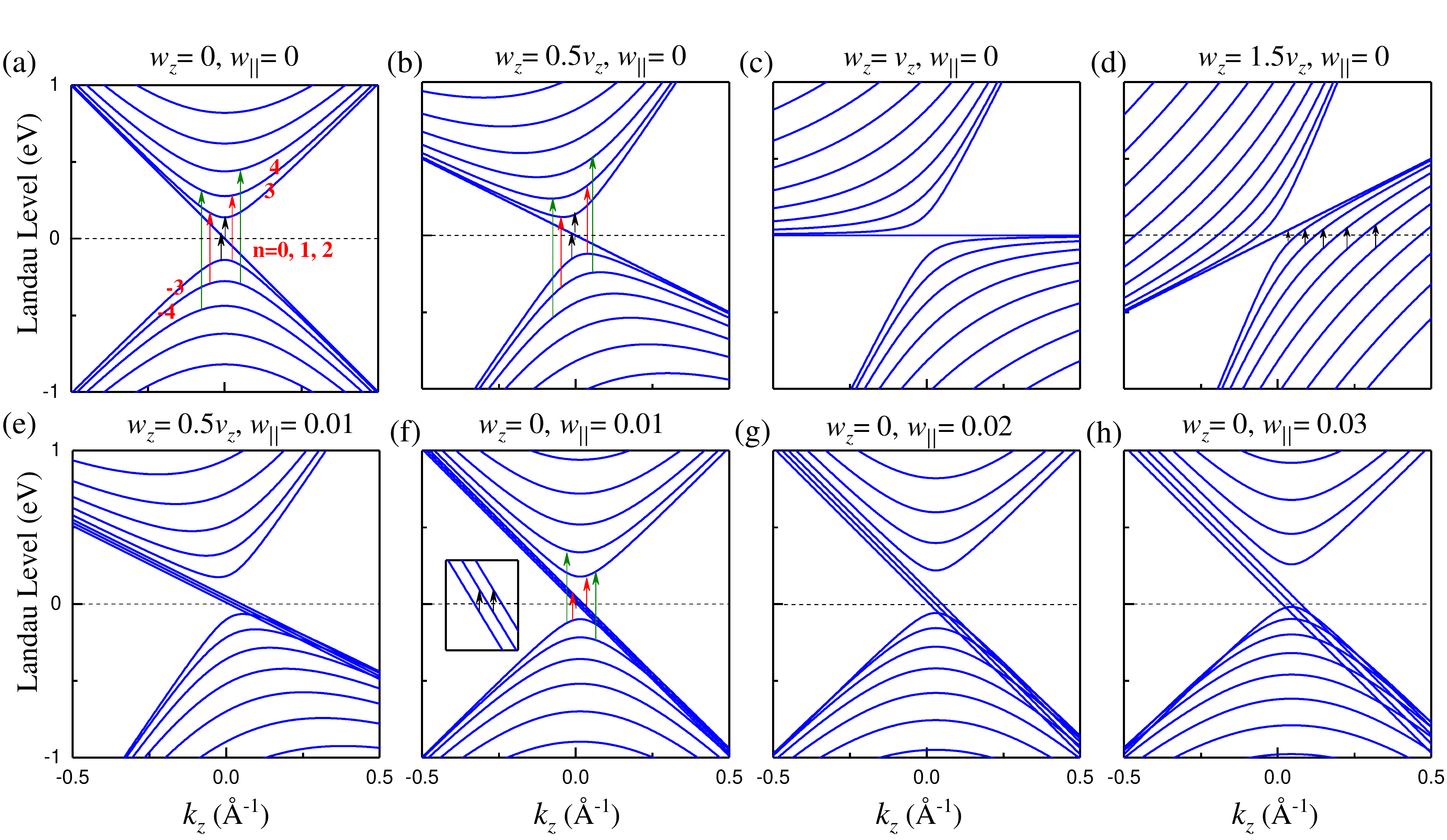}
\caption
{The LL spectrum of the C-3 WPs along $k_z$ with different energy tilts. We set $v_z$=2 eV$\cdot$\AA, $\lambda$=0.02 eV$\cdot$\AA$^3$ and $l_B$=1 \AA. The arrows mark some representative optical transitions. The units of the parameters of the quadratic tilt $w_{||}$ is eV$\cdot$\AA$^2$.}
\label{fig.2}
\end{figure*}

Similarly, the LL spectrum and the eigenstates of C-3 WP ($m=3$) are obtained as
\begin{eqnarray}
\varepsilon_{n}(k_{z}) & = & (2|n|+1)\omega_{c}+(w_{z}-v_{z})k_{z}, \label{eq:LL30}\\
\Psi_{n} & = & (0,\left\vert n\right\rangle)^T,\label{eq:wf30}
\end{eqnarray}
for $n=0,1,2$, and
\begin{eqnarray}
\varepsilon_{n}(k_{z}) & = & 2(|n|-1)\omega_{c}+w_{z}k_{z}\pm\Gamma_{n}^{\prime}(k_{z}),\label{eq:LL31}\\
\Psi_{n} & = & (\alpha'_{n}\left\vert n-3\right\rangle
,\beta'_{n}\left\vert n\right\rangle )^T,\label{eq:wf31}
\end{eqnarray}
for $|n|\geqslant3$ with

\begin{eqnarray}
\Gamma^{\prime}(k_{z}) & = & \sqrt{\left(v_{z}k_{z}-3\omega_{c}\right)^{2}+\frac{8\lambda^{2}}{l_{B}^{6}}|n|(|n|-1)(|n|-2)}, \nonumber \\
\\
\alpha'_n&=&\sqrt{\frac{v_zk_z-3\omega_{c}\pm \Gamma'(k_z)}{\pm2\Gamma'(k_z)}}, \\
\beta'_n&=&\sqrt{\frac{-v_zk_z+3\omega_{c}\pm \Gamma'(k_z)}{\pm2\Gamma'(k_z)}}.
\end{eqnarray}

\begin{figure*}
\includegraphics[width=1.6\columnwidth]{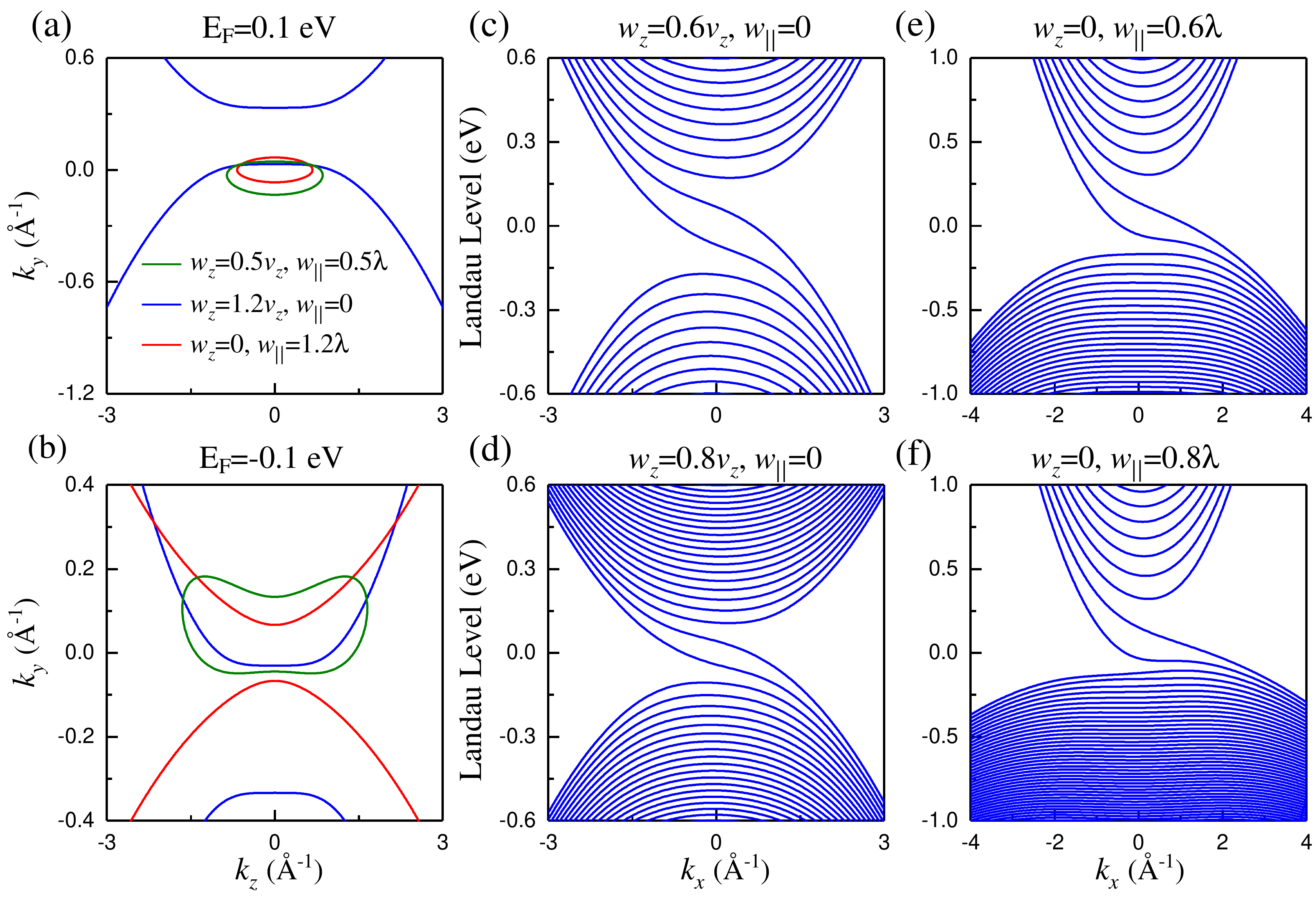}
\caption
{LL spectrum of C-2 WP for $\boldsymbol{B}\parallel x$. (a-b) Semiclassical orbit of Type I (green solid line), II (blue solid line) and III (red solid line) C-2  WPs at $k_x=0$ plane for (a) conduction band  and (b) valence band. (c-f) LL spectrum of C-2 WP with different energy tilts for $l_B$=1 \AA. We set $v_z$=1.5 eV$\cdot$\AA~and $\lambda=0.1$ eV$\cdot$\AA$^2$. }
\label{fig.3}
\end{figure*}

\begin{figure*}
\includegraphics[width=1.6\columnwidth]{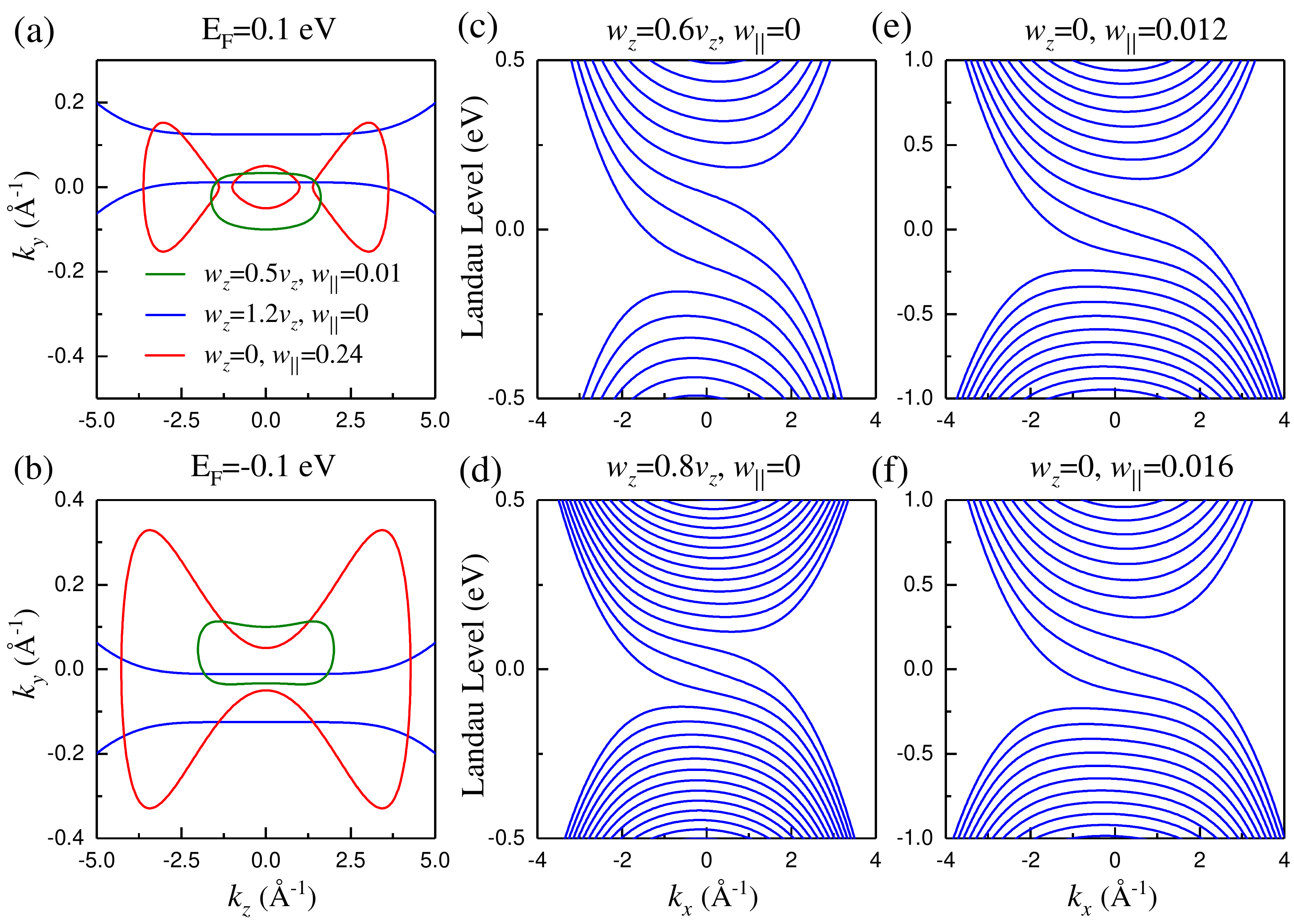}
\caption
{LL spectrum of C-3 WP for $\boldsymbol{B}\parallel x$. (a-b) Semiclassical orbit of Type I (green solid line), II (blue solid line) and III (red solid line) C-3  WPs at $k_x=0$ plane for (a) conduction band and (b) valence band. (c-f) LL spectrum of C-3 WP with different energy tilts for $l_B$=1 \AA. We set $v_z$=2 eV$\cdot$\AA~and $\lambda$=0.02 eV$\cdot$\AA$^3$. The units of the parameters of the quadratic tilt ($w_{||}$) is eV$\cdot$\AA$^2$. }
\label{fig.4}
\end{figure*}

\begin{figure*}
\includegraphics[width=1.6\columnwidth]{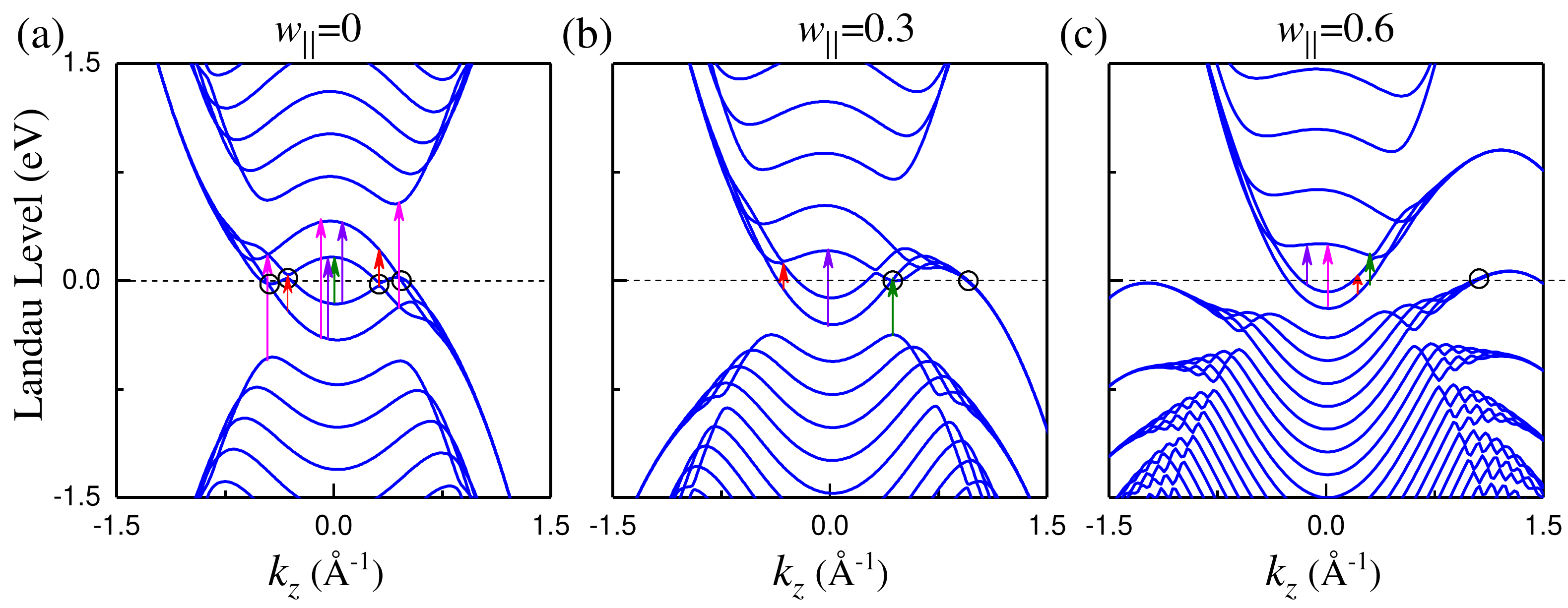}
\caption
{The LL spectrum of the C-4 WP along $k_z$ with  different quadratic energy tilt.
We set $c_1=1$ eV$\cdot$\AA$^{3}$, $c_2=1$ eV$\cdot$\AA$^{2}$ and $l_B$=3.16 \AA. The unit of $w_{||}$ is eV$\cdot$\AA$^2$. }
\label{fig.5}
\end{figure*}

The calculated LLs of C-2 and C-3 WPs with different parameters are
shown in Fig.~\ref{fig.1} and Fig.~\ref{fig.2}. From Eq. (\ref{eq:LL21}) and (\ref{eq:LL31}), one observes that in $k_{z}=0$
plane, the LL energy of C-2 WP is proportional to the magnetic field
$B$ whether there is a tilt term or not, but that of C-3 WP is proportional
to $B^{3/2}$ when $w_{\parallel}=0$, which is different from the
C-1 WP \cite{Marcus}. Besides, we find that the influence of linear
and quadratic energy tilt on LLs are completely different. First,
as shown in the Fig.~\ref{fig.1}(a-d) and~\ref{fig.2}(a-d), without any tilt or  with only
linear tilt, there are two (three) degenerated chiral LLs for C-2
(C-3) WP, consistent with previous results \cite{YongSun,YangGao}. But when
the quadratic energy tilt is added, the degeneracy of the two (three)
chiral bands is broken, as shown in the Fig.~\ref{fig.1}(f-h) [Fig.~\ref{fig.2}(f-h)].
And, one observes that the splitting of the chiral LLs is more significant for C-2 WP.
Moreover, we find that the quadratic energy tilt {[}$(2|n|\pm1)w_{||}/l_{B}^{2}${]} always tends to elevate the LL bands.
Particularly, for  type-III  C-2 WP  $(|w_{||}|>|\lambda|)$, all the LLs at $k_{z}=0$ point have positive energy [see Fig. \ref{fig.1}(h)]. But, this is
not the case for type-III C-3 WP, due to the presence of $k_{+(-)}^{3}$
terms in Hamiltonian (Eq. \ref{eq:ham1}). Second, the linear energy tilt tends to
tilt and squeeze the LL spectrum of C-2 (C-3) WP, while no obvious
tilt of the LLs can be observed for the C-2 (C-3) WP with only quadratic
energy. Increasing $w_{z}$ (with $w_{\parallel}=0$), the LLs gets
more and more tilted, and the chiral LLs become flat bands when $w_{z}=v_{z}$,
as the slope of the chiral LLs is determined by $w_{z}-v_{z}$ {[}see
Eq. (\ref{eq:LL20}) and (\ref{eq:LL30}){]}. When $|w_{z}|>|v_{z}|$, the system becomes type-II
WPs, and the resulting LLs are overtilted. In contrast, increasing
$w_{||}$ ($w_{||}>0$ with $w_{z}=0$), the LLs of the valence band will be squeezed while that of the conduction band are broadened [see Fig.~\ref{fig.1}(f-h) and Fig.~\ref{fig.2}(f-h)].
Particularly, for C-2 WP, the valence band at $k_{z}=0$ plane becomes a flat band when $w_{||}=\lambda$. Generally, the LLs of a trivial flat band are
all degenerate at same energy. But, for a nontrival band with finite
orbital momentum, its structure will be modified when a magnetic field
is applied \cite{ZhiMingYu2,QNiu,QNiu2,QNiu3}. Since the C-2 WP exhibit significant
Berry curvature and orbital momentum, the original flat band here
is modified to have an energy variation, leading to non-degenerate
LLs at $k_{z}=0$ plane, as shown in Fig. \ref{fig.1}(g). Interestingly, J.-W. Rhim et. al. show that
in such case, the LL spacing is determined by the  quantum distance of the original Bloch states \cite{BJYang}. When $|w_{\parallel}|>|\lambda|$ in C-2 WP and $|w_{\parallel}|>0$ in C-3 WP, the systems become type-III WPs, and the Fermi surface is no longer a point but contains two hole pockets, and then the Fermi level may or may not  go though many LLs depending on the size of  the two hole pockets and the strength of $B$ field  {[}see Fig.~\ref{fig.1}(h) and~\ref{fig.2}(h){]}. When both linear and
quadratic energy tilt are added, not only does the band tilt, but
also the degeneracy of the chiral LLs is broken, as shown in the Fig.~\ref{fig.1}(e) and~\ref{fig.2}(e).

When the $B$ field deviates from $z$ direction, it is generally
difficult to obtain analytical expressions for the LL spectrum. Alternatively,
we calculate it numerically. It was shown that the type-II C-1 WP
features LL collapse which can be quantitatively understood within
a semiclassical picture \cite{ZhiMingYu2}. Under a magnetic field, the semiclassical
orbit of electronic motion in momentum space resides on the intersection
between a constant energy surface and a plane perpendicular to the
field direction. By tuning $\boldsymbol{B}$ away from linear tilt direction ($z$-direction), the semiclassical orbit of type-II C-1 WP changes from
a circle to an open trajectory, leading to the collapse of LLs \cite{ZhiMingYu2}.
In this work, we show that when $B$ field deviates from $z$ direction, the type-II C-2 and C-3 WPs also feature the LL collapse for both conduction and valence bands. But for type-III C-2 WP, only the LL spectrum of valence (conduction) band collapses when $w_{\parallel}>0$ ($w_{\parallel}<0$) and  the type-III C-3 WP generally does not feature LL collapse.

To explicitly show this, we plot the constant energy surface of the C-2 and C-3 WPs with $k_{x}=0$,
i.e. we assume $B$ field along $x$ direction, for different linear
and quadratic energy tilt in Fig.~\ref{fig.3}(a-b) and~\ref{fig.4}(a-b). One can find that for type-I
(type-II) C-2 and C-3 WPs, the constant energy surfaces are always
closed (open).
In sharp contrast, for type-III C-2 WP, the constant energy surface of  conduction band is still closed but that of valence band
becomes an open trajectory [see the red line in Fig. \ref{fig.3}(a-b)]. And no open constant energy surface can
be found in type-III C-3 WP [see the red line in Fig. \ref{fig.4}(a-b)], as the $\pm\lambda k_{\parallel}^{3}$
dominates the band structure for large $k_{\parallel}$.

We also numerically calculate the LLs of C-2 and C-3 WPs for $\boldsymbol{B}\parallel x$
in the region before collapse. The results show that when approaching
the phase boundary between type-I and type-II, the LL spacing becomes
more and more small, as shown in the Fig.~\ref{fig.3}(c-d) and~\ref{fig.4}(c-d). When the LL spacing vanishes, the LLs collapse.
However, the LL spacing of the valence (conduction) band of C-2 WP becomes more and more small (large) when increasing the quadratic tilt ($w_{||}>0$), as shown in Fig.~\ref{fig.3}(e-f).
Fig. \ref{fig.4}(e-f) show that  while the LL spacing of  C-3 WP also varies with $w_{\parallel}$, it generally does not approach zero, and hence the  LLs will not collapse, consistent with the semiclassical analysis.

\subsection{C-4 WP}

Different from the C-2 and C-3 WPs, which can appear at high-symmetry
points and lines, the C-4 WP only appear at the high-symmetry points
with ${\cal T}$or ${\cal T}^{\prime}$ symmetry \cite{ZMYu,GuiBinLiu,ZeyingZhang}. Hence, the C-4 WP
only has quadratic energy tilt, and a general form of the effective
Hamiltonian of C-4 is \cite{ZMYu}
\begin{eqnarray}
{\cal H}_{C-4}&=& w_{||}k^{2}+c_{1}k_{x}k_{y}k_{z}\sigma_{y}\nonumber \\
 &  & +c_{2}[\sqrt{3}(k_{x}^{2}-k_{y}^{2})\sigma_{x}+(k_{x}^{2}+k_{y}^{2}-2k_{z}^{2})\sigma_{z}],\label{eq:hamc-4}
\end{eqnarray}
where $k=\sqrt{k_{x}^{2}+k_{y}^{2}+k_{z}^{2}}$, $w_{||}$ denotes the
strength of quadratic energy tilt and $c_{1(2)}$ is real model parameters.
The energy dispersion of C-4 WP reads
\begin{eqnarray}\label{eq: hamc-4E}
E_{\pm}&=&w_{||}k^2 \pm \sqrt{4c_2^2k^4-12c_2^2G(k)+c_1^2k_x^2k_y^2k_z^2},\nonumber \\
\end{eqnarray}
where $G(k)=k_x^2k_y^2+k_y^2k_z^2+k_z^2k_x^2$. Because of the complexity of Eq. (\ref{eq:hamc-4}), it is difficult
to solve the LLs and eigenvector of C-4 WPs analytically. Hence, in
following calculations, we numerically calculate  the LL spectrum of C-4 WP with different quadratic energy tilt.
As shown in the Fig.~\ref{fig.5}(a), without any tilt, there are four non-degenerated chiral LLs crossing the zero energy.
By adding the quadratic energy tilt, the both chiral and achiral LL bands are elevated, but the elevation is more significant for the chiral LLs in $k_z>0$ region.
Moreover, we find that increasing $w_{||}>0$, the LLs of valence band will be squeezed while that of conduction band are broadened.

\begin{figure}
\includegraphics[width=1\columnwidth]{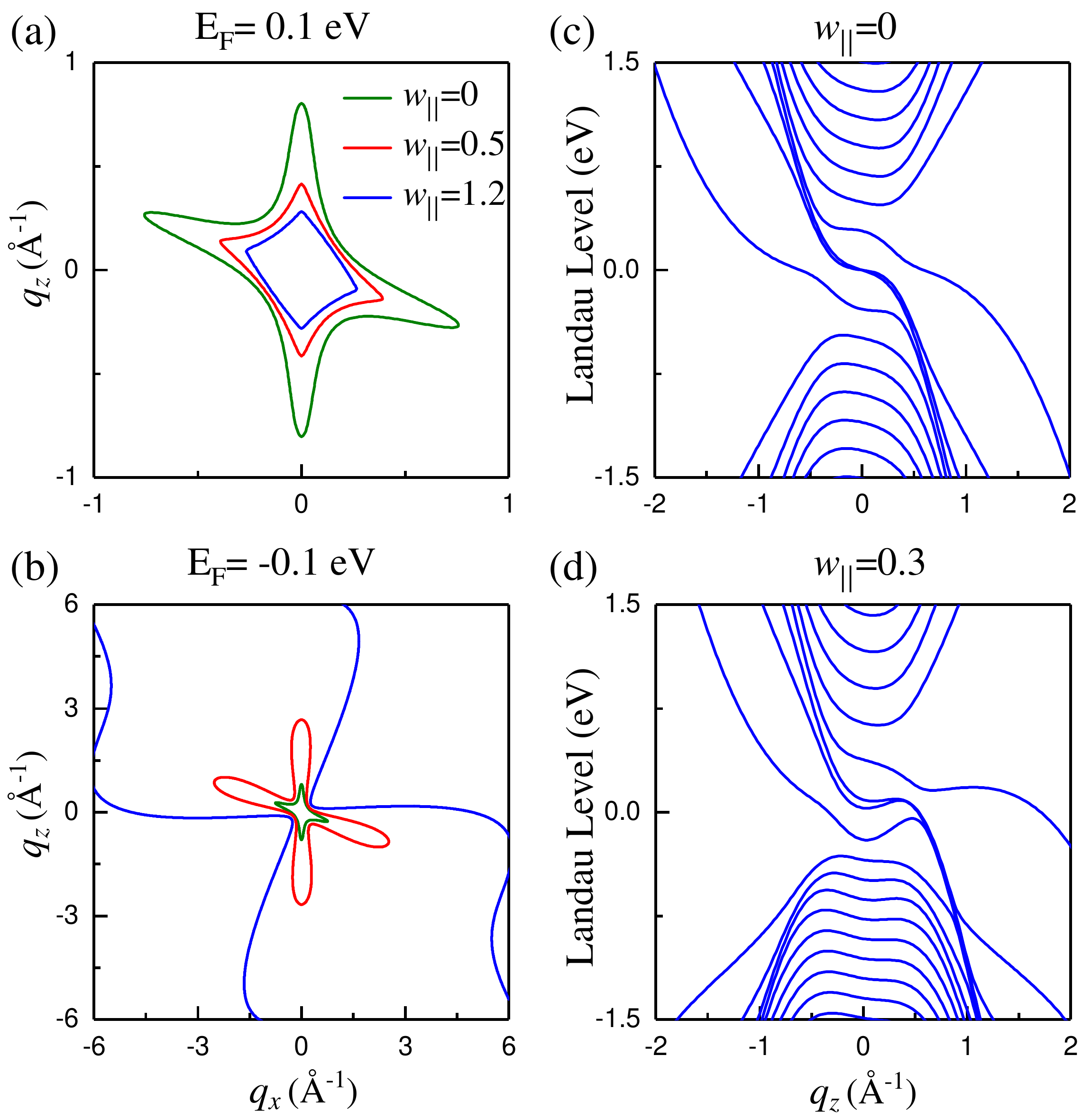}
\caption
{LL spectrum of C-4 WP for $\boldsymbol{B}$ $\parallel (111)$ direction. Semiclassical orbit of C-4 WPs at $q_y=0$ plane with different quadratic energy tilts for (a) conduction band and (b) valence band.
(c-d) LL spectrum of C-4 WP with different parameters for $l_B$=3.16 \AA.  We set $c_1=1$ eV$\cdot$\AA$^{3}$ and $c_2=1$ eV$\cdot$\AA$^{2}$. }
\label{fig.6}
\end{figure}

Again, we study the possibility of LL collapse in C-4 WP.  From Eq.  (\ref{eq: hamc-4E}), one knows that the C-4 WP exhibits a leading order of cubic along the (111) direction (as well as the directions symmetry-connected to it), and of quadratic  along other directions. Hence, when the energy tilt $w_{||}$ is large enough and tuning $\boldsymbol{B}$ away from $z$-direction, the semiclassical orbit of C-4 WP will go from closed to open, leading to the collapse of LLs. To explicitly show it,  we rotate the coordinate axis of system,  making  $k_{x,y,z}$ to be $q_{x,y,z}$. Here,  $q_z$ is along the original (111) direction.
We plot the constant energy surface of the C-4 WP with $q_y=0$, i.e, we assume $B$ field along the (111) direction, for different quadratic energy tilt in Fig. \ref{fig.6}(a-b). One can find that  when $w_{||}$ is large ($w_{||}>0$), the constant energy surfaces of conduction (valence) band are always closed (open). We also numerically calculate the LLs of C-4 WPs for $B$ along (111) direction before collapse. The results show that the LL spacing of the valence (conduction) band of C-4 WP becomes more and more small (large) when increasing the quadratic tilt $w_{||}>0$, as shown in the Fig.~\ref{fig.6}(c-d). The spacing of LL for valence (conduction) band approach zero and the LL spectrum of valence (conduction) band collapses when the $w_{||}$ is large enough ($w_{||}>0$).

\begin{figure*}
\includegraphics[width=16 cm]{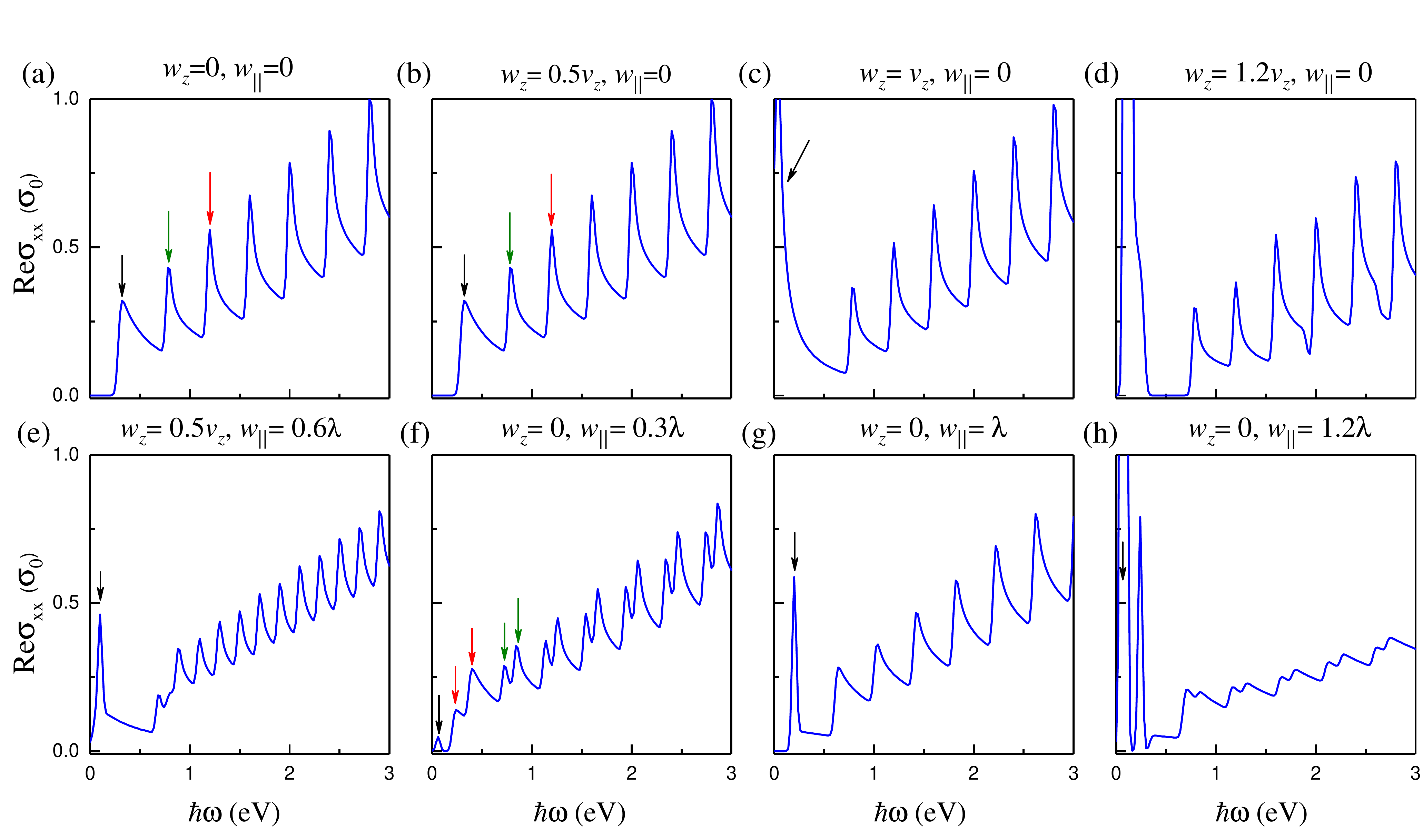}
\caption
{Re[$\sigma_{xx}(\omega)$] plotted for the C-2 WPs with linear and quadratic tilt terms, in units of $\sigma_0=\frac{\hbar e^2}{2\pi l_B^2}$,  and the used parameters are the same as those in Fig.~\ref{fig.1}.  The black, red and green arrows correspond to the transitions between the different subbands in Fig.~\ref{fig.1}.}
\label{fig.7}
\end{figure*}

\section{Magneto-optical conductivity \label{sec:III}}
In this section,  we study the magneto-optical conductivity of the unconventional WPs. The dynamical conductivity tensor can be obtained from the Kubo formula \cite{Tsuneya,GDMahan}. Here, we  focus on the absorptive part of the longitudinal magneto-optical conductivity with a $x$-linearly polarized light and $\boldsymbol{B}\parallel z$.
Expressed in the Landau level basis in the clean limit, we have \cite{Tsuneya,GDMahan}
\begin{eqnarray}
\text{Re}[\sigma_{xx}(\omega)]= & \frac{\hbar e^{2}}{2\pi l_{B}^{2}}\sum_{n,n'}\int dk_{z}\frac{\Delta f|\langle n,k_{z}|v_{x}|n',k_{z}\rangle|^{2}}{\Delta\varepsilon_{n}(\varepsilon_{n}-\varepsilon_{n'}+\hbar\omega+i\Gamma)},
\end{eqnarray}
where $|n,k_{z}\rangle$ denotes the LL state, $v_{x}=\partial_{k_{x}}{\cal H}$
is the velocity operator, $\hbar\omega$ is the photon energy, $\Delta\varepsilon_{n}=\varepsilon_{n}-\varepsilon_{n'}$
and $\Delta f=f_n-f_{n'}$ are the energy and the occupation differences between
the two states involved in the optical transition and $f(E)=1/[e^{(E-E_{F})/k_{B}T}+1]$
is the Fermi-Dirac distribution function with Boltzman constant $k_{B}$
and temperature $T$.

\subsection{Optical transition selection rules}

For C-2 and C-3 WPs, the velocity operator is given by
\begin{eqnarray}
v_{x}&=&\frac{\partial{\cal H}}{\partial k_{x}}=\omega_1(\hat{a}^{\dagger}+\hat{a})+\omega_2\left[\hat{a}^{m-1}\sigma_{+}+\left(\hat{a}^{\dagger}\right)^{m-1}\sigma_{-}\right],\nonumber \\\label{eq:vx}
\end{eqnarray}
where $\omega_1$=$\sqrt{2}w_{||}/l_{B}$ and $\omega_2$=$m\lambda(\sqrt{2}/{ l_B})^{m-1}$. Substituting the eigenvector of C-2 Eq. (\ref{eq:wf20}) and (\ref{eq:wf21}) into Eq. (\ref{eq:vx}), we can
get $v_{n,n'}=\langle n,k_{z}|v_{x}|n',k_{z}\rangle$ of the form
\begin{eqnarray}
v_{n,n^{\prime}} & =&[\omega_1\alpha_n^*\alpha_{n+1} \sqrt{n-1}
+\omega_1\beta_n^*\beta_{n+1} \sqrt{n+1} \nonumber \\
&&+\omega_2\beta_n^*\alpha_{n+1}\sqrt{n}]+h.c.,
\end{eqnarray}
from which the optical transition selection rules can be inferred, \emph{i.e.} the optical transitions only happen between the $|n|$-th and $|n|\pm1$-th
LLs.

Similarly, the optical transition matrix elements $v_{n,n'}$ of C-3 WP is obtained as,
\begin{eqnarray}
v_{n,n^{\prime}} & =&[\omega_1\alpha_n^*\alpha_{n+1} \sqrt{n-2}
+\omega_1\beta_n^*\beta_{n+1} \sqrt{n+1} \nonumber \\
&&+\omega_2\beta_n^*\alpha_{n+1}\sqrt{n(n+1)}]+h.c.,
\end{eqnarray}
Again, the optical transition only happen between the $|n|$-th and $|n|\pm1$-th LLs. Actually, the dipolar transitions existing in C-2 and C-3 WPs results from an emergent infinity rotation symmetry along $z$-direction in the effective Hamiltonian of C-2 and C-3 WPs {[}see Eq. (\ref{eq:ham1}){]}.

Different from the C-2 and C-3 WPs, the C-4 WP does not have infinity
rotation symmetry along any direction, hence both dipolar and non-dipolar
transitions can be found in  C-4 WP.

\subsection{Numerical results }

\begin{figure*}
\includegraphics[width= 16 cm]{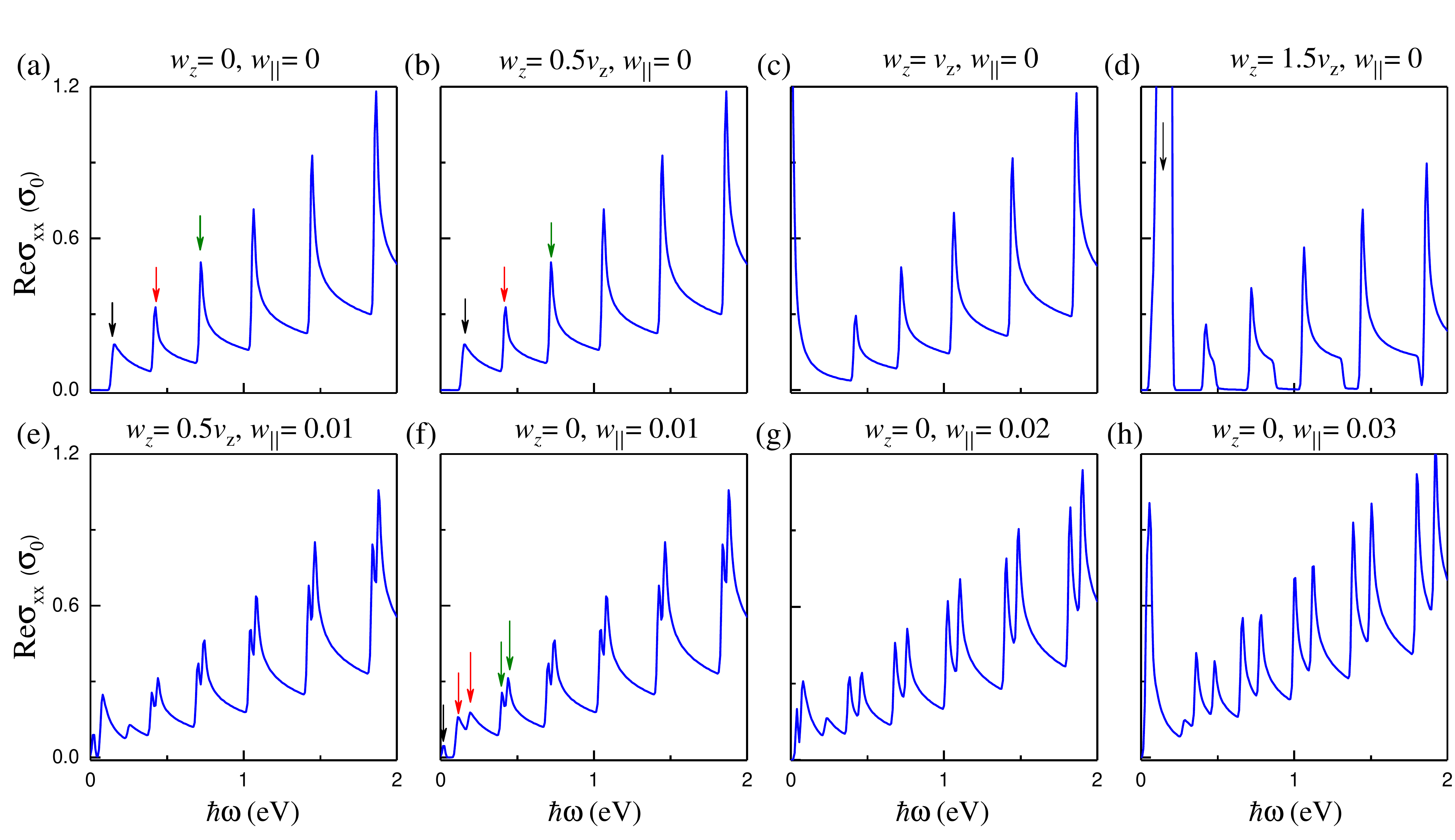}
\caption
{Re[$\sigma_{xx}(\omega)$] plotted for the C-3 WPs with linear and quadratic tilt terms, in units of $\sigma_0=\frac{\hbar e^2}{2\pi l_B^2}$,  and the used parameters are the same as those in Fig.~\ref{fig.2}.  The black, red and green arrows correspond to the transitions between the different subbands in Fig.~\ref{fig.2}.}
\label{fig.8}
\end{figure*}

We then investigate the influence of  the linear and quadratic energy tilt on the  magneto-optical measurements. The calculated results
of $\text{Re}[\sigma_{xx}(\omega)]$ of C-2 and C-3 WPs are  plotted
in Fig.~\ref{fig.7} and Fig.~\ref{fig.8}, and the used parameters are the same as
those in Fig.~\ref{fig.1} and Fig.~\ref{fig.2}, respectively. By inspecting the LL
spectra in Fig.~\ref{fig.1} and Fig.~\ref{fig.2}, we can observe the following features.

First, without any energy tilt, the conductivity includes a series of asymmetric peaks and the peak spacing is proportional to $B$ ($B^{3/2}$)
for C-2 (C-3) WPs {[}see Fig.~\ref{fig.7}(a) and~\ref{fig.8}(a){]}.
This  can be directly obtained by checking the LL spectrum {[}Eq. (\ref{eq:LL21}) and (\ref{eq:LL31}){]}, where the energy
extremal points reside  at $k_{z}=0$  point {[}see Fig.~\ref{fig.1}(a) and~\ref{fig.2}(a){]}. The positions of the peaks can be easily obtaining by analyzing the transitions happening between the LLs in $k_{z}=0$ plane.

Second,  $\text{Re}[\sigma_{xx}(\omega)]$ does not have a dependence on linear energy tilt in type-I phase {[}see Fig.~\ref{fig.7}(a-b) and~\ref{fig.8}(a-b){]}, but shows distinctive features in type-II phase. Since the Weyl cone is overtilted in type-II
phase and the system becomes a metal rather than a semimetal, the
Fermi level will cross both chiral and achiral LLs, and in such case,
many transitions that can (not) happen in type-I phase are forbidden
(occur), as shown in Fig.~\ref{fig.7}(d) and~\ref{fig.8}(d). As a consequence, many absorption peaks
appear at low frequencies, and the shape of the original peaks are
changed from exhibiting long tails to tailless. These are similar to the case in C-1 WP \cite{ZhiMingYu2}.

Third, as aforementioned, the quadratic energy tilt breaks the degeneracy
of the two (three) chiral LLs of C-2 (C-3) WP. This leads to two interesting
results. One is the transition between chiral LLs, giving rise to one additional peak at low frequencies, as showing in Fig.~\ref{fig.7}(f) and Fig.~\ref{fig.8}(f), respectively (marked by the black arrows). Remarkably, we find that the position of these
additional peaks is $2w_{||}/l_{B}^{2}$, indicating that the peak
position can be tuned by the strength of \textbf{$B$} field and the
quadratic energy tilt. The other interesting result is that the energy
degeneracy of the transitions involving chiral LLs also is broken.
For C-2 WP with $w_{||}=0$, the transition energy for $-2\rightarrow 1$
and that for $1\rightarrow 2$ are the same, and for C-3 WP $-3\rightarrow 2$ and $2\rightarrow 3$ are
the same, which marked by red arrow in Fig.~\ref{fig.1}(a) and Fig.~\ref{fig.2}(a).  In the presence of finite $w_{||}$, these degeneracies are broken and then the original peaks in conductivity are splitted, as shown in Fig.~\ref{fig.7}(f) and~\ref{fig.8}(f) (marked by red arrows).
In addition,  the quadratic energy tilt also modifies the energy of  achiral LLs of both C-2 and C-3 WPs, making the transition energy between $-|n|\rightarrow |n|-1$
and $-(|n|-1)\rightarrow |n|$ are no longer the same.
Thus the peaks resulting from transitions between the achiral bands also undergoes splitting, as indicated by the green arrows in Fig.~\ref{fig.1}(f) and Fig.~\ref{fig.7}(f), and that in Fig.~\ref{fig.2}(f) and Fig.~\ref{fig.8}(f).

\begin{figure*}
\includegraphics[width=1.6\columnwidth]{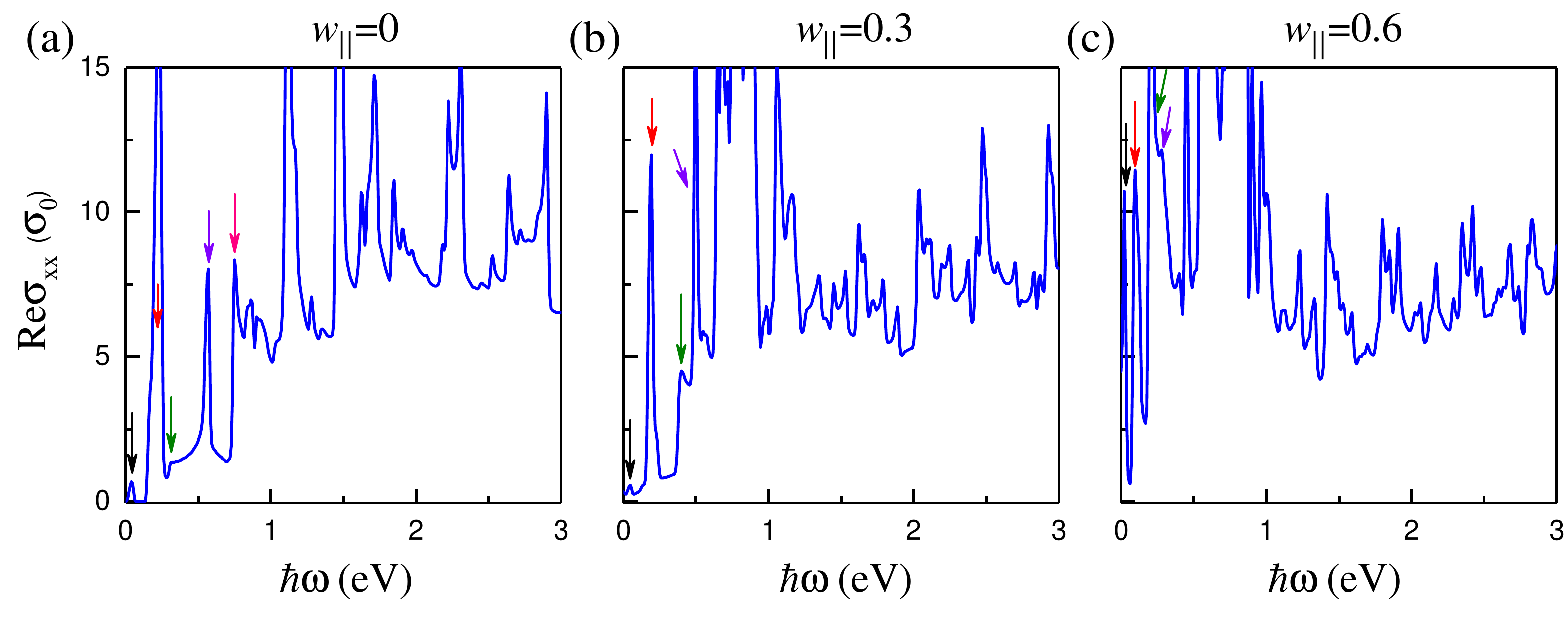}
\caption
{Re[$\sigma_{xx}(\omega)$] plotted for the C-4 WPs with quadratic tilt terms, in units of $\sigma_0=\frac{\hbar e^2}{2\pi l_B^2}$,  and the used parameters are the same as those in Fig.~\ref{fig.5}.  The red and green arrows correspond to the transitions between the different subbands in Fig.~\ref{fig.5}. The black arrows correspond to the transitions that occur at the $k$ point marked by the black circle in Fig.~\ref{fig.5}.}
\label{fig.9}
\end{figure*}

Fourth, for type-III C-2 WP, there also exist invariable presence
of absorption peaks at low frequencies due to the metal feature [see Fig.~\ref{fig.7}(h)]. But,
for type-III C-3 WPs, the LLs of conduction and valence bands generally
are separated in energy, when the strength of $B$ field is large or $w_\parallel$ is small. Hence, the type-III C-3 WP may or may not
exhibit significant absorption peaks at low frequencies, depending
on the model parameters, as shown in Fig.~\ref{fig.8}(f-h).

At last, we discuss the magneto-optical conductivity of C-4 WP, the results of which are plotted in Fig.~\ref{fig.9}.
One observes that the magneto-optical conductivity of C-4 WP is irregular due to the complexity of the corresponding LLs, completely differing  from that of C-2 and C-3 WPs.
Thus, the influence of quadratic energy tilt on the magneto-optical conductivity also is subtle and may have a strong dependence on the model parameters and the strength of $B$ field.
But two key observations can be inferred from the LLs and Fig.~\ref{fig.9}.
Since the four chiral LLs in C-4 WP are always not degenerate, the C-4 WP inevitably exhibits several transition peaks at low-energy, as labelled by the arrows in Fig.~\ref{fig.9}.
Besides, there is almost a absorption peak around the zero-energy position, as the Fermi level always crosses the intersection of two subbands, as marked by the black circles in Fig.~\ref{fig.5}.

\section{Summary\label{sec:IV}}

In summary, based on the low-energy effective Hamiltonian and Kubo formula, we have systematically studied the magnetoresponse of WPs with both linear and quadratic energy tilt, and with a topological charge of $n=2,3,4$. We find that the magnetoresponses of these systems with linear and that with quadratic energy tilt exhibit completely different signatures. For C-2 and C-3 WPs, the linear energy tilt always tends to squeeze the LLs of both conduction and valence bands, and eventually leads to LL collapse in type-II phase. In contrast, for C-2 and C-4 WPs, the  quadratic energy tilt tends to squeeze the LLs of either the  valence or the  conduction band, and broadens the LLs of the other one. Hence, when quadratic energy tilt is large and the C-2 (C-4) WP becomes type-III one, only valence or conduction band features LLs collapse. Interestingly, the LL collapse general can not be realized in C-3 WP regardless of the presence or absence of the quadratic energy tilt.
We also find that for type-I WPs, the magneto-optical conductivity does not have a dependence on the linear energy tilt.
But when the linear energy tilt is large enough and the type-I WPs become type-II ones,  many additional absorption peaks appear at low frequencies.
In contrast, the degeneracy of the chiral LLs for C-2 and C-3 WPs can  be broken by the quadratic energy tilt, which makes  new transition between the chiral LLs and lead to one additional peak in the magneto-optical conductivity spectrum at low frequencies.
The modification of LLs bands caused by quadratic tilt also  lead to the splitting of the original magneto-optical conductivity peaks induced by the transitions between achiral LLs.
The transition between different chiral LLs always can be found in C-4 WP, as the four chiral LLs in C-4 WP are not degenerate whether or not there is a quadratic energy tilt.

\section{Acknowledgement}
The authors thank J. Xun for helpful discussions.
This work was supported by the National Key R\&D Program of China (Grant No. 2020YFA0308800), the NSF of China (Grants Nos. 12234003, 12061131002 and 12004035), and the  National Natural Science Fund for Excellent Young Scientists Fund Program (Overseas) and Beijing Institute of Technology Research Fund Program for Young Scholars.

\bibliography{MO_Uncon_Weyl_v1207}

\end{document}